\documentclass[DIV15]{scrartcl}

\usepackage{graphicx}
\usepackage{amsfonts}
\usepackage{amssymb}
\usepackage{amsmath,amsfonts,amsthm,bm}
\usepackage{mathrsfs}
\usepackage{verbatim}
\usepackage{subfigure}
\usepackage{authblk}

\newcommand{\VECE}{\textbf{E}_\text{ext}}
\newcommand{\E}{E_\text{ext}}
\newcommand{\Ebar}{e}
\newcommand{\emax}{u_\text{max}}
\newcommand{\emin}{u_\text{min}}

\newcommand{\VECn}{\textbf{n}}
\newcommand{\VECz}{\textbf{z}}

\begin{document}

\title{Flexoelectric fluid membrane vesicles in spherical confinement}

\author[1]{Niloufar Abtahi}
\author[2]{Lila Bouzar}
\author[2]{Nadia Saidi-Amroun}
\author[3]{Martin Michael M\"{u}ller}

  \affil[1]{\normalsize Department of Physics, Faculty of Arts and Sciences, Eastern Mediterranean University; Famagusta, North Cyprus via Mersin 10, Turkey}
  \affil[2]{\normalsize Laboratoire de Physique des Mat\'eriaux, Universit\'e des Sciences et de la Technologie Houari Boumediene; 
   BP~32 El-Alia Bab-Ezzouar, 16111 Alger, Algeria}
  \affil[3]{\normalsize Laboratoire de Physique et Chimie Th\'eoriques - UMR 7019, Universit\'e de Lorraine; 1 boulevard Arago, 57070 Metz, France}


\date{\vspace{-5ex}}

\maketitle

\begin{abstract}
The morphology of spherically confined flexoelectric fluid membrane vesicles in an external uniform electric field is studied numerically. Due to the deformations induced by the confinement, the membrane becomes  polarized resulting in an interaction with the external field. The equilibrium shapes of the vesicle without electric field can be classified in a geometrical phase diagram as a function of scaled area and reduced volume \cite{Kahraman2012A,Kahraman2012B}. When the area of the membrane is only slightly larger than the area of the confining sphere, a single axisymmetric invagination appears. A non-vanishing electric field induces an additional elongation of the confined vesicle which is either perpendicular or parallel depending on the sign of the electric field parameter. Higher values of the surface area or the electric field parameter can reduce the symmetry of the system leading to more complex folding. We present the resulting shapes and show that transition lines are shifted in the presence of an electric field.
The obtained folding patterns could be of interest for biophysical and technological applications alike.
\end{abstract}



\section{Introduction}
Dielectric materials display a variety of electromechanical coupling mechanisms \cite{Ahmadpoor2015}. A classic example is piezoelectricity, which was discovered by Jacques and Pierre Curie in 1880 \cite{Curie1880}. Piezoelectricity results from a linear coupling allowing a crystalline material with no inversion symmetry to convert a uniformly applied electric field into a mechanical contraction or dilatation and vice-versa. It has been exploited for industrial and biomedical applications such as sensorics \cite{Gautschi2002}, artificial muscles \cite{Madden2004}, or implantology \cite{Labanca2008}. 

In 1969 Robert B. Meyer suggested an analogous linear coupling mechanism between the electric polarization and the curvature strain of liquid crystals \cite{Meyer1969}. 
In contrast to piezoelectricity, flexoelectricity is a ubiquitous phenomenon displayed by all dielectrics.  
In particular, bendable two-dimensional structures like fluid lipid membranes and graphene sheets exhibit the flexoelectric effect \cite{Ahmadpoor2015}. In the context of flexoelectric fluid membranes 
Petrov and coworkers have shown its relevance for biological membranes in experiments \cite{Petrov2001,Petrov2006,Petrov2013}, which motivated several subsequent theoretical studies \cite{Rey2006,Gao2008,Mohammadi2014,Steigmann2016}. Only recently it has been shown with extensive molecular dynamics simulations that uniform electric fields can induce biologically relevant membrane deformations \cite{Khandelia2016}. 
In this theoretical letter, we focus on such a flexoelectric closed fluid membrane, which is confined inside a spherical cavity of smaller size. The whole system is immersed in a uniform electric field. 
The case without electromechanical coupling has been studied in several publications for confined vesicles with spherical topology \cite{Kahraman2012A,Kahraman2012B,Sakashita2014,Purohit2014,Noguchi2019} and higher genus as well \cite{Bouzar2015,Noguchi2015}. To include the flexoelectric effect, we base our model on a recent theory by Steigmann and Agrawal, which was obtained as the thin-film limit of the continuum electrodynamics of nematic liquid crystals \cite{Steigmann2016}.

Using finite element simulations and the `shooting method' for axisymmetric configurations, we determine the shapes of the confined vesicle as a function of surface area, enclosed volume and electric field parameter. In particular when the electric field parameter is large we obtain intricate folding patterns that result from the interplay of the confinement and the flexoelectric effect.


\section{The model}
\noindent
Lipid bilayer membranes consist of two layers of polar lipids which stay in contact with each other due to the amphiphilic structure of the lipids. The thickness of such a bilayer is typically much smaller than its lateral extension, which implies that it can be modelled as a two-dimensional surface $\Omega$. Moreover, since the lipids can move freely within each layer, one can consider the bilayer as fluid in the tangential plane. Including flexoelectricity, one has to take into account four main contributions to describe the mechanics of such a membrane \cite{Kahraman2012B,Steigmann2016}: the bending energy, the surface energy which originates from changes in the membrane's area, the pressure difference between the inner and the outer part of the membrane, and finally, the electromechanical energy which arises from the response of the electrically polarized surface to an external electric field. 

According to the classical spontaneous curvature model, the bending energy $E_\text{b}$ of a vesicle is described by a surface integral involving a second order expansion in curvatures \cite{Canham1970,Helfrich1973,Seifert1997,Steigmann2003}
\begin{equation}
\label{eqn:bendingenergy}
E_\text{b}=\int_{\Omega}
{ \left[\frac{\kappa}{2}(2H-C_{0})^{2}+\overline{\kappa}K_{G}\right]\text{d}A} 
\; ,
\end{equation}
where $H$, $K_{G}$, and $C_{0}$ are the mean, the Gaussian, and the spontaneous curvature, respectively. 
The constant ${\kappa}$ is the bending rigidity  and ${\overline{\kappa}}$ denotes the saddle-splay modulus. 
In the following, we will only consider vesicles of spherical topology and zero spontaneous curvature. This allows to discard all terms but the first one involving the square of the mean curvature. 

Since the energy scales of the surface and pressure contributions are much larger than the bending energy \cite{Seifert1997}, we include these terms as constraints on the total surface area $\bar{A}$ and enclosed volume $\bar{V}$ of the membrane vesicle. To simplify the problem, we scale the two quantities with the corresponding area $A_0$ and volume $V_0$ of the confining container \cite{Kahraman2012A,Kahraman2012B}: 
\begin{equation}
\label{eqn:aandv}
  a=\bar{A}/A_0 \quad \text{and} \quad v=\bar{V}/V_0
  \; .
\end{equation}

Flexoelectricity adds another contribution to the total energy when an external electric field $\VECE$ is present. Following Steigmann and Agrawal the corresponding energy of the flexoelectric membrane is given by \cite{Steigmann2016}:
\begin{equation}
\label{eqn:flexoenergy}
E_\text{f}=\int_{\Omega} { \left[\frac{1}{2D}[(\VECE \cdot\VECn)^{2}-\left|\VECE\right|^{2}]\right]\text{d}A } 
\; ,
\end{equation}
where $\VECn$ is the normal vector of the membrane surface. In this model the authors assume free charges to be absent. Moreover, the polarisation vector is supposed to be essentially tangential to the membrane surface, a simplification which is supported to a certain extent by quantum mechanical considerations and molecular dynamics simulations \cite{Seelig1978,Frischleder1982,Warshaviak2011}. The electric self-field of the membrane can be neglected in this case, which yields a justification for its suppression \cite{Steigmann2016}. 

The material constant $D=\chi_\perp-\frac{c_2^2}{k_3}$ indicates the strength of the flexoelectric effect. It is a combination of the inverse of the electric polarisability, $\chi_\perp$, exhibited by the membrane 
when the electric field $\VECE$ acts in the tangential plane, the bend modulus $k_3$ of the Frank energy of nematic liquid crystals, and the flexoelectric constant $c_2$ in the coupling term between the polarisation vector and the tangent vectors of the membrane \cite{Steigmann2016}. The strength of the flexoelectric effect will determine the sign of $D$. For a constant electric field pointing in the vertical $\VECz$-direction, $\VECE=\E\VECz$, one can define a dimensionless electric field parameter
\begin{equation}
  \label{eqn:Ebar}
  \Ebar = \frac{\E^2 R^2}{D \kappa} \; ,
\end{equation}
where $R$ is the radius of the confining sphere.  
This allows writing the scaled total energy of a free flexoelectric membrane vesicle as:
\begin{equation}
\label{eqn:totalenergy}
\tilde{E}_\text{tot} =  \int_{\Omega} u \, \text{d}\tilde{A}  
=  \int_{\Omega} \left[ 2\tilde{H}^2 + \frac{1}{2}\Ebar [(\VECz \cdot\VECn)^{2}-1]\right]\text{d}\tilde{A}  \, + \, \text{constraints}
\; ,
\end{equation}
where all lengths are scaled with $R$.
The expressions for the constraints on area and volume depend on the numerical solution method as explained below. Inverting the direction of the electric field does not change the equilibrium shapes, since $\VECE$ enters the equations quadratically.

To include the effect of the confinement, we model the spherical cavity as a rigid container without adhesion between the membrane and the cavity.
A simple calculation shows that a uniform electric field, which is applied at infinity, stays uniform inside the container as long as the whole system---except the membrane vesicle which does not contribute due to the neglection of the self-field---consists of isotropic dielectric media \cite{Jackson1999}. This rather crude approximation allows to obtain a first idea of the shapes that flexoelectric membrane vesicles can adopt in a spherical confinement. To be closer to more realistic experimental setups, one would have to take into account an electrolyte in the interior of the container and add an adhesion energy for the contact between membrane and container.


\section{Numerical solution methods}
Equilibrium solutions are determined by minimising Eq.~(\ref{eqn:totalenergy}) for fixed parameters $a$, $v$, and $\Ebar$. In this letter we use two different numerical methods: $(i)$ Axisymmetric shapes without self-contact can be determined with the help of a Hamiltonian formulation. The resulting differential equations are solved with a classical shooting method. The container is treated as a hard constraint. Area and volume are conserved with the help of Lagrange multipliers. With this method we search for the simplest axisymmetric shapes consisting of one free part and one part in contact with the container. 
$(ii)$ The second approach is based on the finite element method and can account for more complicated shapes including self-contacts and symmetry breaking. The surface of the membrane is discretized into a triangular mesh. A discretized version of Eq.~(\ref{eqn:totalenergy}) is used to determine the forces that act on each node. To equilibrate the system, we add a damping force on the nodes and integrate Newton's equations of motion in time. In the simulations the container is modelled as a soft constraint with a quadratic repulsive force. The constraints on area and volume are implemented \text{via} a penalty method. 
More details on both methods can be found in the supplementary material.


\section{Axisymmetric solutions}

\begin{figure}
\centering
\subfigure[][]{\raisebox{0.5cm}{\label{fig:axisymmetrya}\includegraphics[width=0.4\textwidth]{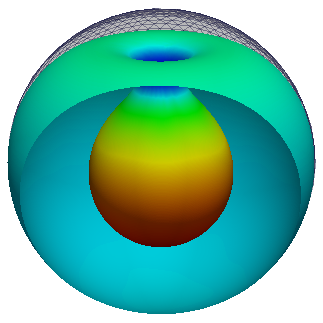}}}
\raisebox{0.0cm}{\includegraphics[width=0.09\textwidth]{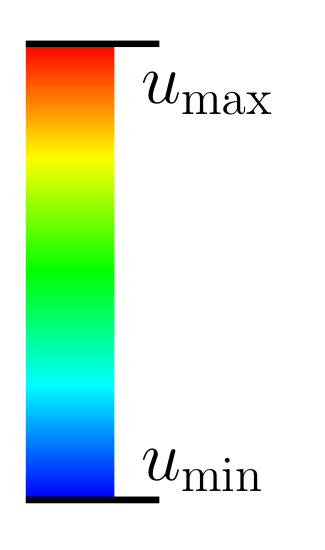}}
\subfigure[][]{\label{fig:axisymmetryb}\includegraphics[width=0.48\textwidth]{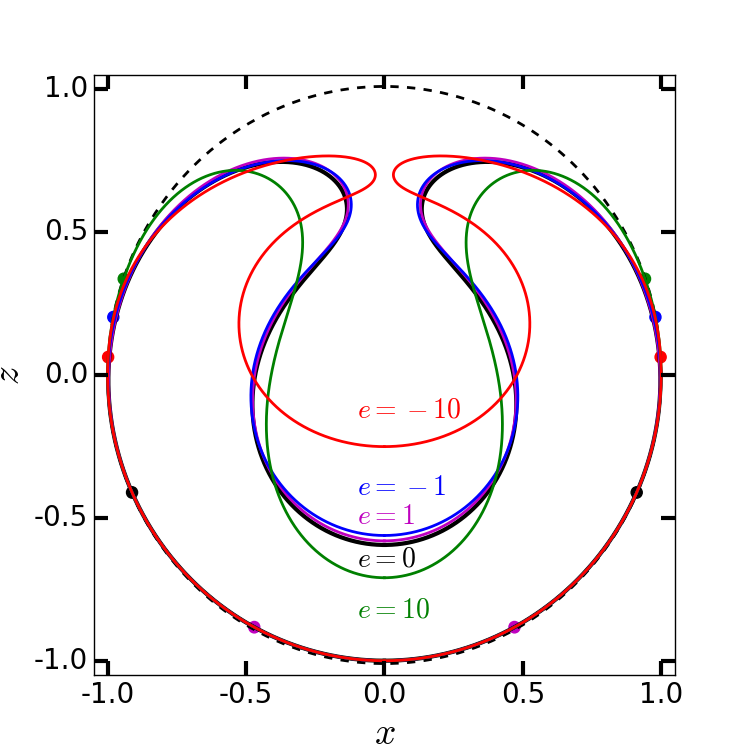}}
\\
\subfigure[][]{\label{fig:axisymmetryc}\includegraphics[width=0.4\textwidth]{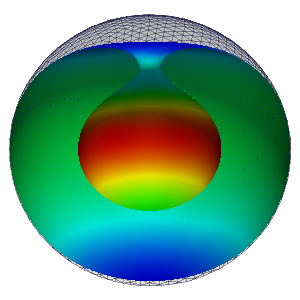}}
\hfill
\raisebox{2cm}{\includegraphics[width=0.08\columnwidth]{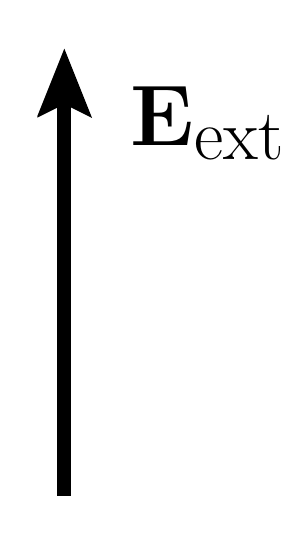}}
\hfill
\subfigure[][]{\label{fig:axisymmetryd}\includegraphics[width=0.4\textwidth]{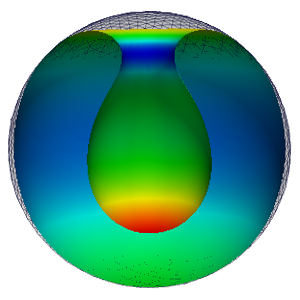}}
\caption {
Numerical equilibrium solutions of a spherically confined flexoelectric membrane vesicle in an external uniform electric field with area $a=1.2$ and volume $v=0.8$ for different values of the electric field parameter $\Ebar$. 
(a) Solution of a finite element simulation for $e=0$ \cite{Kahraman2012A,Kahraman2012B}. 
(b) Axisymmetric solutions obtained with the shooting method for $e=0,\pm 1$ and $\pm 10$. The membrane is 
composed of two segments, one in contact with the confinement and one which is free. The detachment points are indicated with dots. The solutions for $e=0$ and $e=\pm 1$ coincide with the results from finite element simulations within the numerical error. For $e=\pm 10$ one observes an additional detachment of the membrane in the finite element simulations (see Figs.~\ref{fig:axisymmetryc} and \ref{fig:axisymmetryd}), which is not taken into account in the shooting method.
(c) Solution of a finite element simulation for $e=-10$.
(d) Solution of a finite element simulation for $e=10$. 
Colours in (a), (c), and (d) indicate the values of the scaled total energy density ranging from $\emin$ to $\emax$ with $(\emin,\emax)=(0,10)$ in (a), $(2,14)$ in (c), and $(-5,10)$ in (d). The external electric field in (b)-(d) is oriented along the symmetry axis.
}
\label{fig:axisymmetry}
\end{figure}

When the area of the vesicle is larger than the area of the spherical confinement, the membrane has to form an invagination inside the container. This is only possible when the volume enclosed by the vesicle is smaller than the confining volume. For moderate values of the parameters $a$, $v$, and $\Ebar$, the equilibrium solutions consist of an axisymmetric invagination connected to the contacting part of the membrane \textit{via} a neck (see Fig.~\ref{fig:axisymmetry}). To get an idea of the behaviour of the system, we first consider different values of $\Ebar$ with area and volume fixed to $(a,v)=(1.2,0.8)$, which yields the shapes shown in Fig.~\ref{fig:axisymmetry}. 
The \textit{free} vesicle without electric field adopts the form of an oblate-discocyte as the global energy minimum \cite{Seifert1991} \footnote{For the free vesicle with $e=0$ one also finds two local minima with higher bending energy: a stomatocyte and a prolate configuration. The energies of stomatocyte and oblate-discocyte are comparably close. However, the energy barrier between these shapes is high enough to avoid a transition due to thermal energy \cite{Seifert1991}.}. 
Fig.~\ref{fig:axisymmetrya} shows the \textit{confined} equilibrium solution which resembles a stomatocyte \cite{Kahraman2012A,Kahraman2012B}.
The flexoelectric effect induces an elongation of the invagination. Depending on the sign of $\Ebar$ this deformation is either parallel or perpendicular to the direction $\VECz$ of the external electric field  (see Figs.~\ref{fig:axisymmetryb}-\ref{fig:axisymmetryd}) \footnote{When the electric field is not directed along the invagination at the beginning of a simulation, the mesh reorients during equilibration until symmetry axis and electric field coincide.}.

The sign of the electric field parameter, Eq.~(\ref{eqn:Ebar}), depends on the sign of $D$ which, in turn, is determined by the strength of the flexoelectric effect. The latter is encoded in the constant $c_2$. When the flexoelectric effect is weak, $c_2^2 < k_3 \chi_\perp$, and $\Ebar$ is positive, whereas $\Ebar$ is negative for $c_2^2 > k_3 \chi_\perp$. An inspection of Eq.~(\ref{eqn:totalenergy}) reveals that the flexoelectric energy density is minimised when $e (\VECz\cdot\VECn)^2$ is as small as possible. For positive $\Ebar$ this term is minimised when the surface normal $\VECn$ is perpendicular to $\VECz$ leading to an elongation in the direction of the electric field \footnote{Interestingly, this effect can be so dominant that the ground state of the \textit{free} vesicle with positive $e$ is a stomatocyte and not an oblate-discocyte.}. 
When the flexoelectric effect is strong, the surface normal prefers to be parallel to $\VECz$ as far as possible. This explains the elongation of the invagination perpendicular to $\VECE$ for $\Ebar<0$.

Fig.~\ref{fig:axisymmetryb} shows axisymmetric shapes for $e=0,\pm 1$ and $\pm 10$ which were obtained with the shooting method. An electric field parameter of the order of $\pm 1$ does not influence the resulting shape dramatically. One can observe, however, that the circle at which the vesicle detaches from the container depends crucially on the value of $e$ (dots in Fig.~\ref{fig:axisymmetryb}).  
A word of caution is due here. 
For larger values of $|e|$ one observes an additional detachment of the membrane in contact with the container in the finite element simulations (see Figs.~\ref{fig:axisymmetryc} and \ref{fig:axisymmetryd}), 
which we cannot capture with our simple shooting method. The corresponding solutions thus have to be treated with prudence and should always be confirmed by finite element simulations.

\begin{figure}
\centering
\subfigure[][]{\label{fig:phasediagrama}\includegraphics[width=0.49\textwidth]{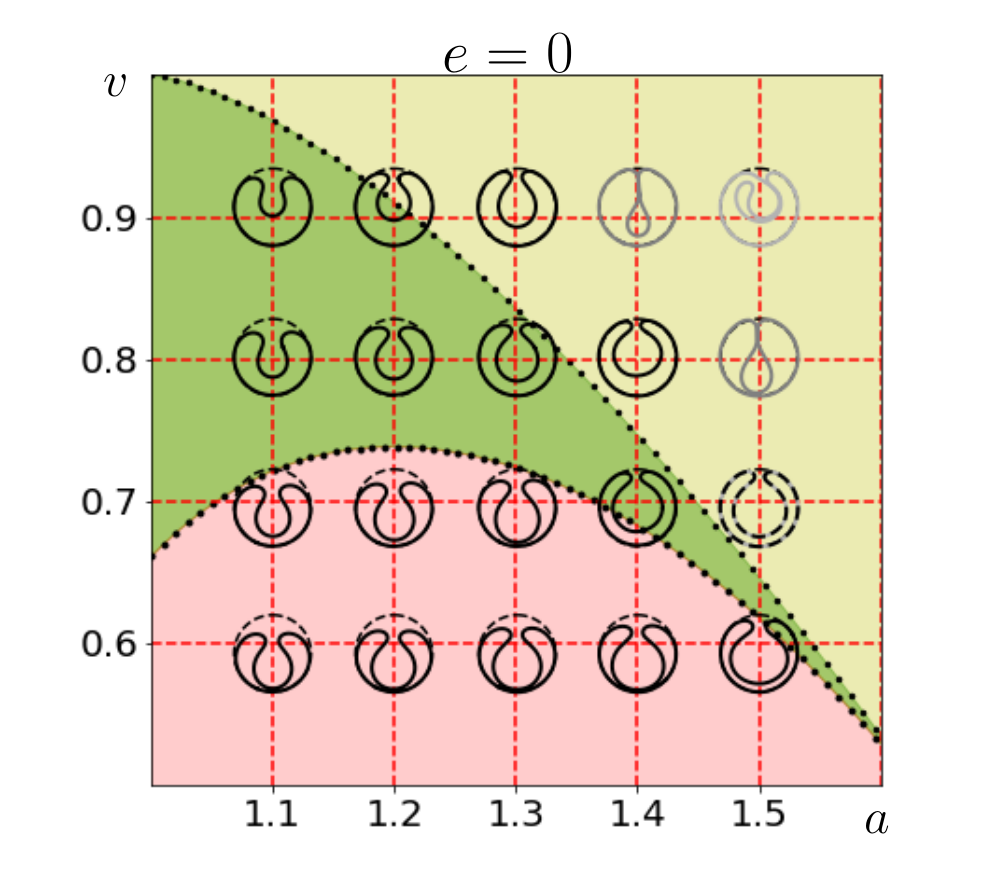}}
\subfigure[][]{\label{fig:phasediagramb}\includegraphics[width=0.49\textwidth]{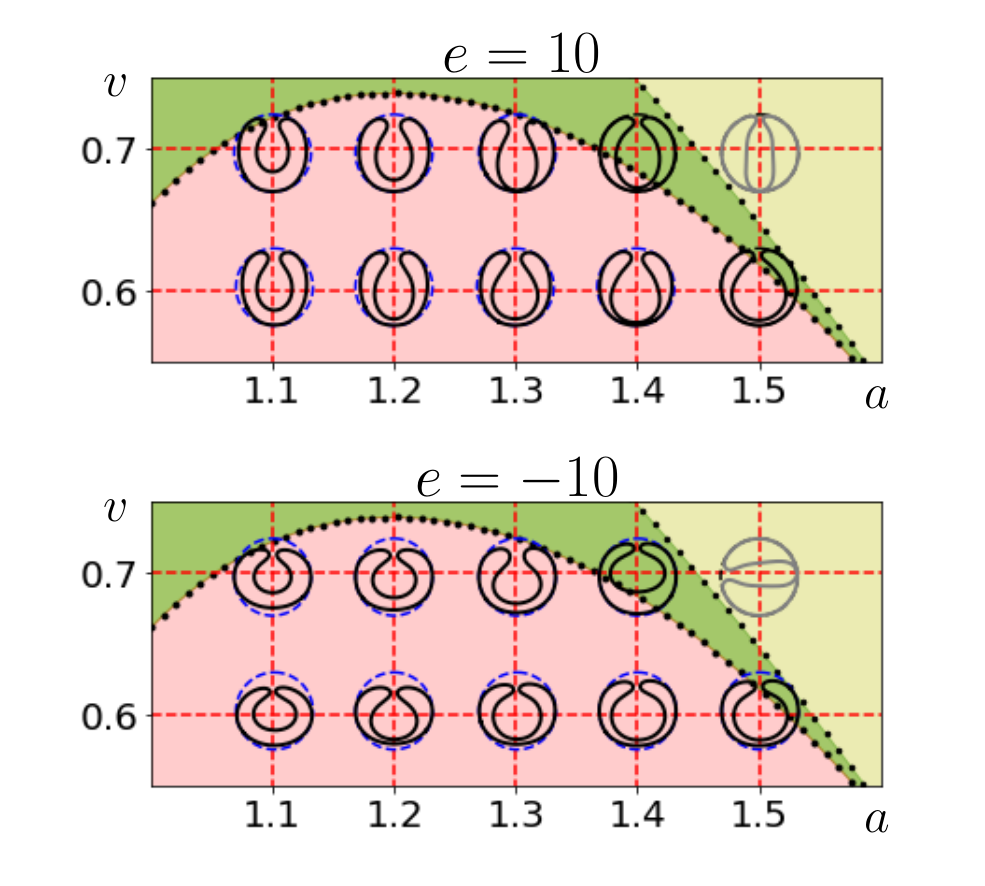}}
\caption {
Morphological phase diagrams with vertical slices of the shapes from finite element simulations (a) for $\Ebar=0$ as obtained in Ref.~\cite{Kahraman2012A} and (b) for $\Ebar=\pm 10$. The dotted lines and coloured regions in all figures were obtained for $\Ebar=0$. In (b) they are duplicated for a better comparison with the detached vesicle shapes (see also main text).}
\label{fig:phasediagram}
\end{figure}

With this in mind we can now take a look at Fig.~\ref{fig:phasediagram}, which displays morphological phase diagrams of our system. 
Fig.~\ref{fig:phasediagrama} recalls the results for the system without electric field, which was studied in 
Refs.~\cite{Kahraman2012A,Kahraman2012B}. The vertical slices were obtained from finite element simulations. In the green region one finds axisymmetric solutions with the shooting method.  
In the pink region below, the solutions are also axisymmetric  
but display a more complicated configuration with several free parts or self-contact, for example. Fig.~\ref{fig:phasediagramb} presents the corresponding part of the phase diagram for $\Ebar=\pm 10$, where the  above-mentioned detachment of the membrane takes place (slices of detached shapes highlighted with confining circle in blue).


\section{Strong electric field and symmetry breaking }

A membrane vesicle in a spherical confinement without electric field $(\Ebar=0)$ can be forced to break axisymmetry by increasing its area above a critical value (corresponding approximately to the upper black curve in Fig.~\ref{fig:phasediagram}) \cite{Kahraman2012A}. Consider, for instance, the vertical slices in Fig.~\ref{fig:phasediagrama} at constant volume $v=0.7$. Below $a=1.5$ the equilibrium solutions are axisymmetric. For $a=1.5$ a metastable ellipsoidal state is observed. The corresponding ground state breaks axisymmetry with an invagination which is reminiscent of a prolate (comparable to the one of 
$(a,v)=(1.5,0.8)$ in Fig.~\ref{fig:phasediagrama}). For $\Ebar=\pm 10$ similar shapes are observed which orient themselves parallel ($e>0$) or perpendicular ($e<0$) to the electric field (see again Fig.~\ref{fig:phasediagramb}). However, the neck connecting the invagination with the rest of the vesicle is not a slit with self-contact but exhibits an ellipsoidal cross section.   

\begin{figure}
\centering
\subfigure[][]{\includegraphics[width=0.44\columnwidth]{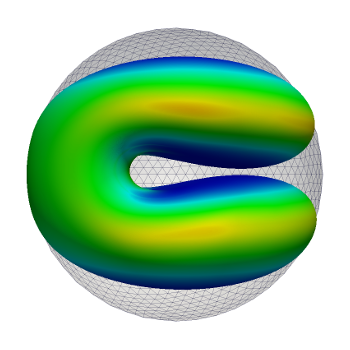}}
\hfill
\raisebox{2cm}{\includegraphics[width=0.08\columnwidth]{efielddirection}}
\hfill
\subfigure[][]{\includegraphics[width=0.44\columnwidth]{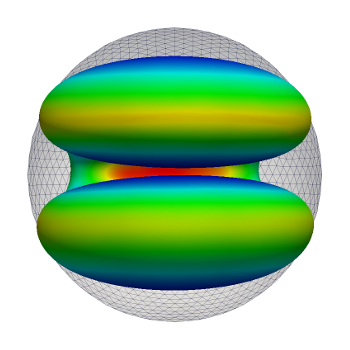}}
\caption {
Symmetry breaking for $a=1.1$, $v=0.7$ and $\Ebar=-50$ (from finite element simulations). The invagination of the membrane is not axisymmetric but deforms into a large elongated slit reminiscent of shapes that can be found with the ADE model for confined membranes without electric field (see Ref.~\cite{Noguchi2019} and references therein). 
(a) Side view and (b) top view of the invagination. The electric field is oriented vertically in (a) and (b). 
The color code of the profiles is the same as in Fig.~\ref{fig:axisymmetry} with 
$(\emin,\emax) = (0,54)$.
}
\label{fig:constantarea}
\end{figure}
A stronger coupling between the membrane and an external electric field provokes 
symmetry breaking at lower values of $a$. 
Fig.~\ref{fig:constantarea} shows one example for small area $a=1.1$ and high \textit{negative} electric field parameter $e=-50$. 
Similar shapes can be found for other areas and volumes as long as the absolute value of $e<0$ is large enough. 
The invagination is slit-like and orients itself perpendicular to the electric field. The minimisation of the flexoelectric energy density, $e (\VECz\cdot\VECn)^2$, now leads to almost flat membrane parts since this term dominates the bending energy term for high $|\Ebar|$.

\begin{figure}[t!]
\centering
\subfigure[][]{\label{fig:constantvolumea}
\begin{minipage}{0.29\textwidth}
\includegraphics[width=\columnwidth]{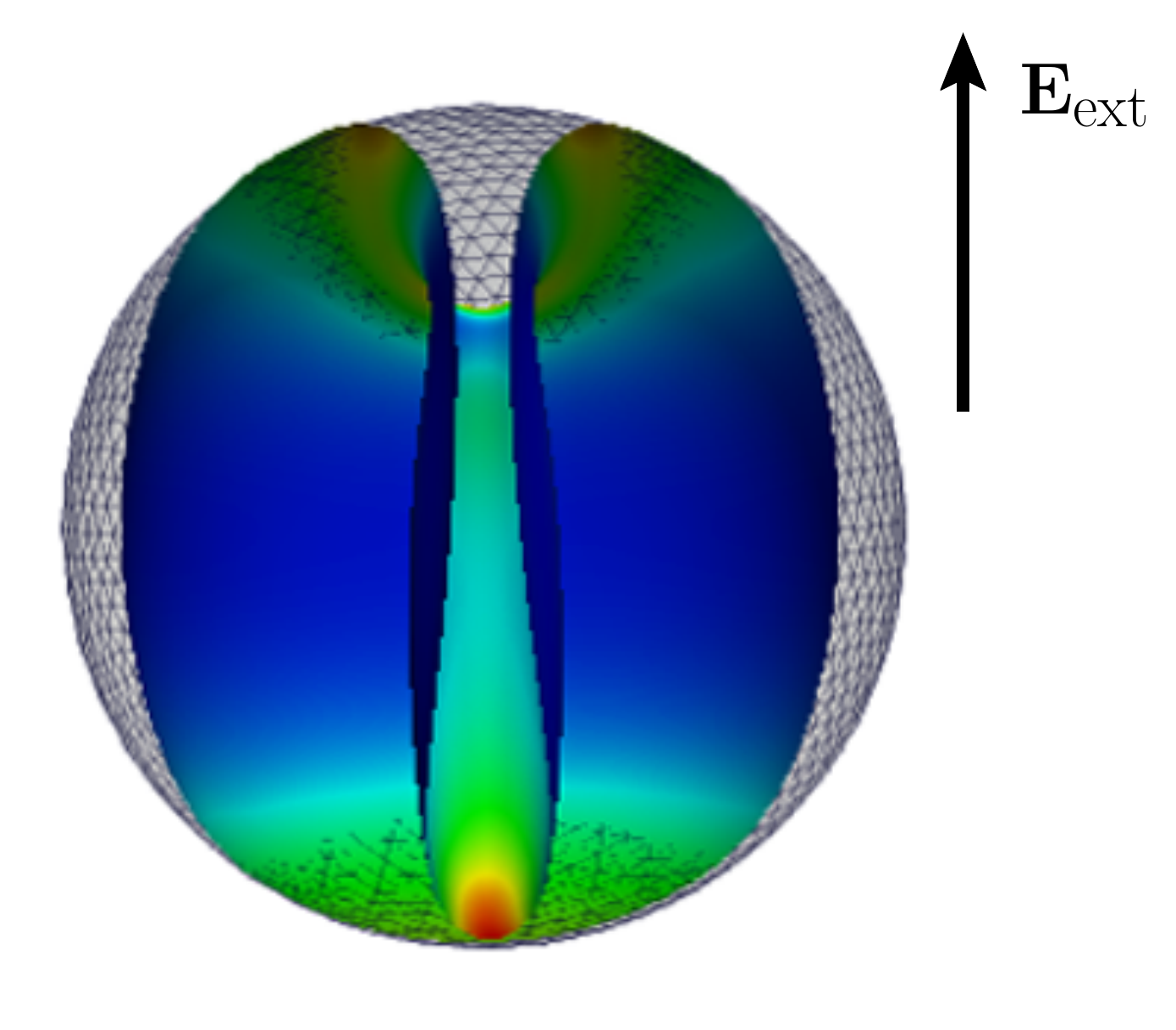}
\\[0.5cm]
\includegraphics[width=\columnwidth]{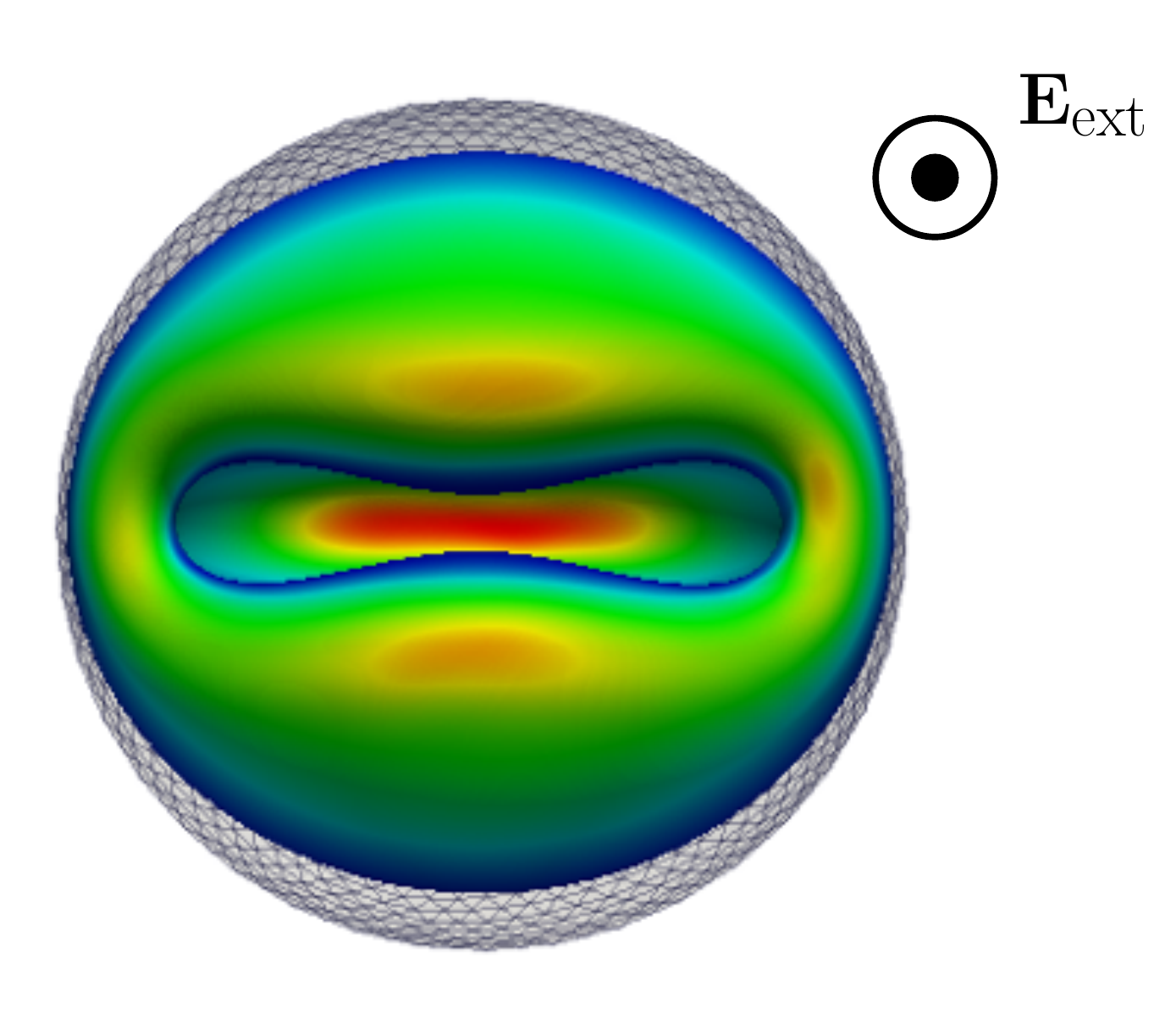}
\end{minipage}
}
\hfill
\subfigure[][]{\label{fig:constantvolumeb}
\begin{minipage}{0.29\textwidth}
\includegraphics[width=\columnwidth]{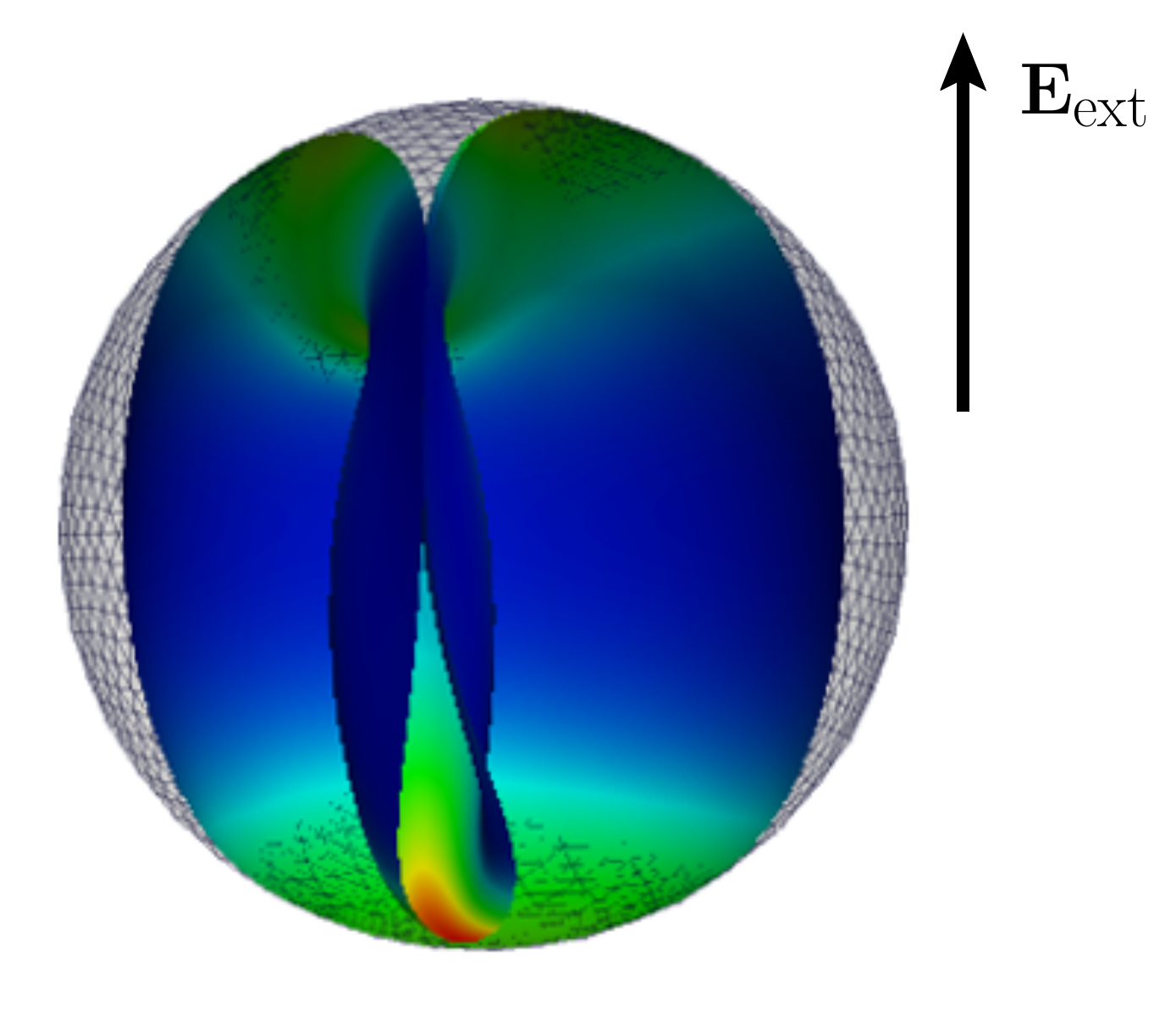}
\\[0.5cm]
\includegraphics[width=\columnwidth]{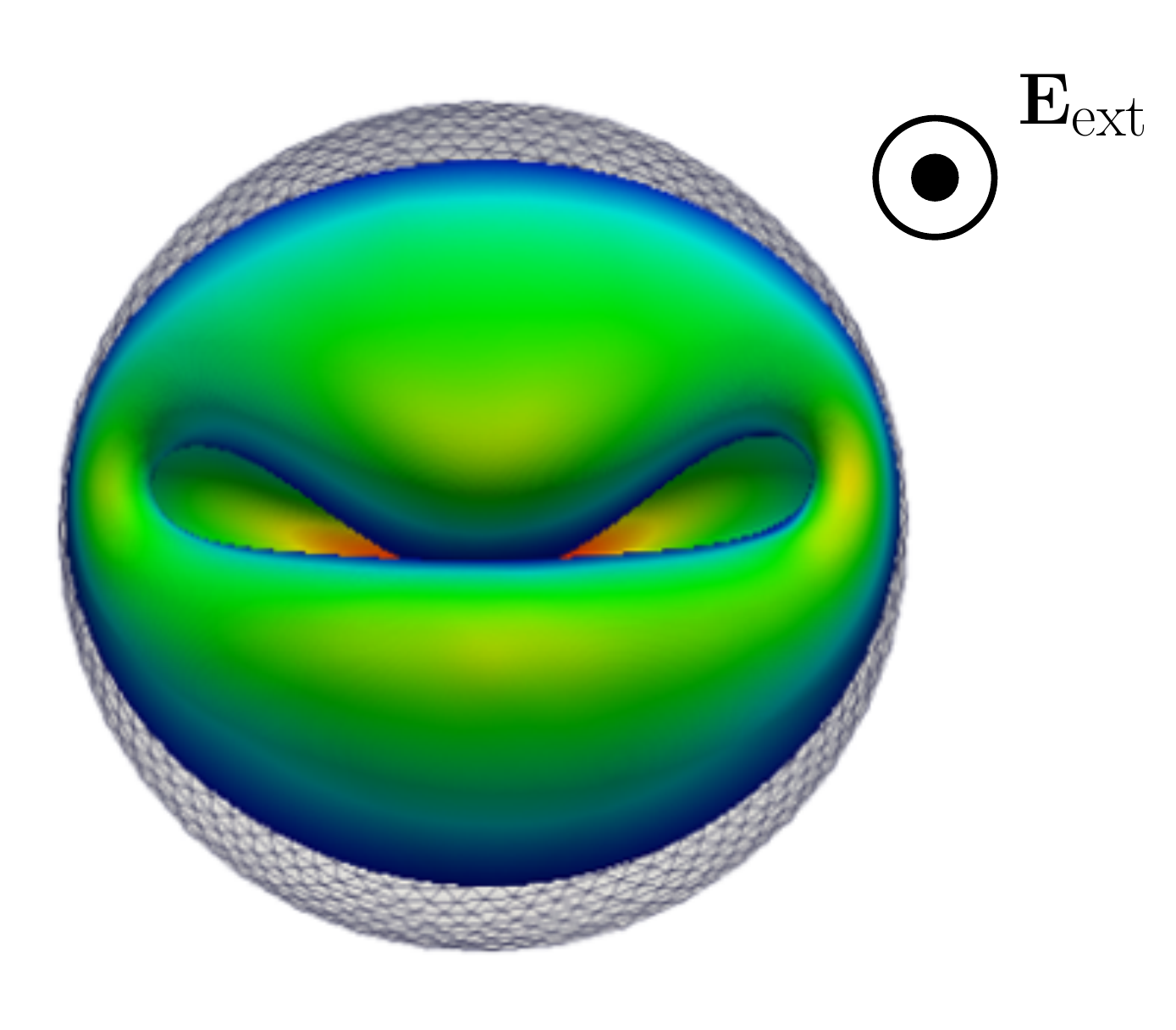}
\end{minipage}
}
\hfill
\subfigure[][]{\label{fig:constantvolumec}
\begin{minipage}{0.29\textwidth}
\includegraphics[width=\columnwidth]{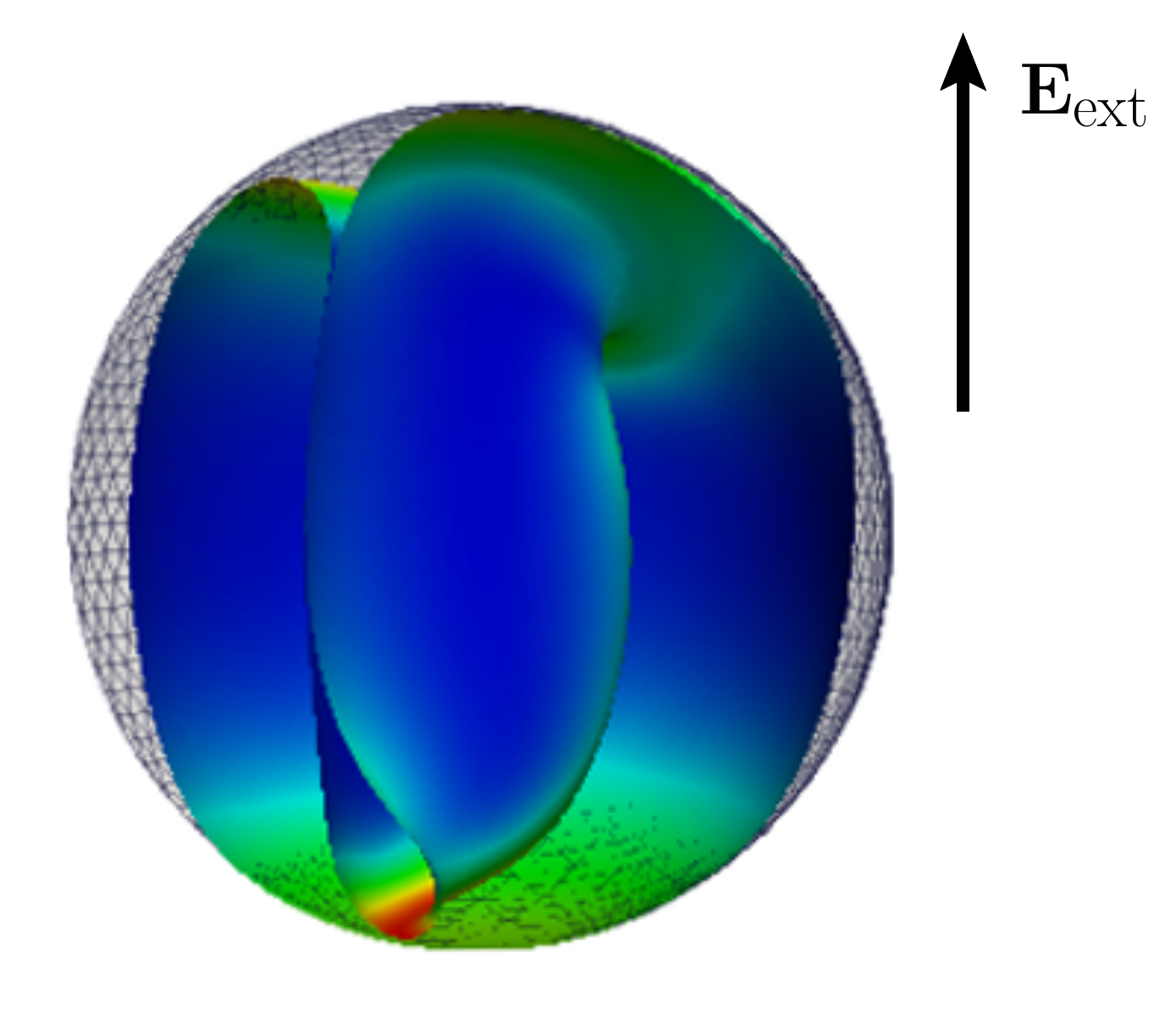}
\\[0.5cm]
\includegraphics[width=\columnwidth]{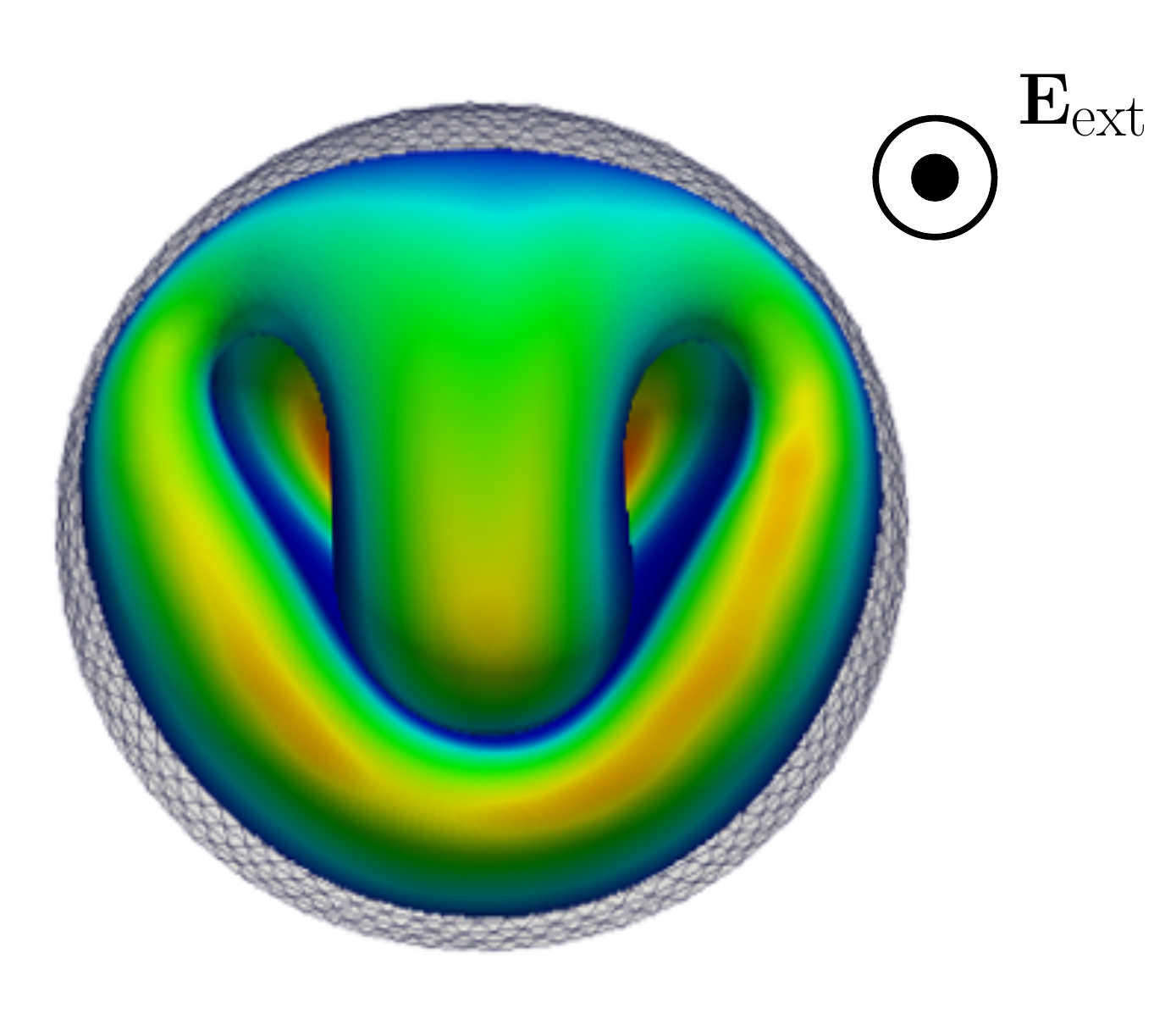}
\end{minipage}
}
\caption {
Symmetry breaking for constant volume $v=0.7$, electric field parameter $\Ebar=100$ and 
increasing area. Equilibrium solutions of finite element simulations with (a) $a=1.3$, (b) $a=1.4$, and (c) $a=1.5$.
(\textit{top}) Side view and (\textit{bottom}) top view of the invagination. The direction of the electric field is indicated in every figure.
The color code of the profiles is the same as in Fig.~\ref{fig:axisymmetry} with $(\emin,\emax) = (-50,30)$ in (a), $(-50,38)$ in (b), and $(-50,31)$ in (c).
}
\label{fig:constantvolume}
\end{figure}

A high \textit{positive} electric field parameter has a similar effect on the membrane vesicle. Fig.~\ref{fig:constantvolume} shows how the electromechanical coupling induces a symmetry breaking for $\Ebar=100$ and constant volume $v=0.7$. The transition is similar to the axisymmetric-to-prolate transition of $\Ebar=0$ but happens already between $a=1.2$ and $1.3$. Fig.~\ref{fig:constantvolumea} shows the ground state for $a=1.3$. The system contains two symmetry planes. The slit-like invagination is parallel to $\VECE$ as expected. For $a=1.4$ large portions of the membrane come into contact. One of the planar symmetries is broken (see Fig.~\ref{fig:constantvolumeb}). A further increase of $a$ deforms the slit-like neck even further (see Fig.~\ref{fig:constantvolumec}). 


\section{Conclusions}
In this letter we have studied how a confined flexoelectric fluid membrane vesicle responds to an external electric field. To find equilibrium configurations as a function of area, volume and the coupling with the electric field, two numerical solutions methods were exploited. Despite some rather crude approximations such as assuming a constant electric field, we have found exciting shape transformations and symmetry breaking. 

Self-contacts as observed in this work can potentially lead to a transition from a spherical to a toroidal vesicle topology \textit{via} membrane fusion. Some of the autors studied this question for the system without electric field \cite{Bouzar2015}. It turns out that the spherical topology is preferred for typical values of the material parameters. Flexoelectricity could potentially facilitate topology changes. For a definite statement, however, one would have to study confined flexoelectric fluid membrane vesicles of toroidal topology in detail for $\Ebar\neq 0$ which goes beyond the scope of this paper.

To confirm that the obtained shapes are stable and do not rupture due to the stresses induced by the electric field in experiments, one has to check whether the resulting surface tension is below the membrane's rupture tension. This question can be addressed qualitatively by comparing the membrane stresses due to the area constraint and the external electric field. The latter is linear in the electric field parameter $e$ whereas the former is linear in the 
scaled surface tension $\tilde{\sigma}=\frac{\sigma R^2}{\kappa}$, where $\sigma$ is the unscaled surface tension and $R$ is the radius of the confining sphere (see supplementary material). 
One expects a maximum surface tension of the order of a few mN/m, which is approximately the rupture tension for a fluid phospholipid bilayer \cite{Evans2003}. Inserting typical values for the other parameters,  $R=0.5\,\mu$m and $\kappa=20 k_B T$, one obtains $\tilde{\sigma}\approx 3000$ which is much larger than the values of $e$ that we consider.

Experiments on confined fluid membrane vesicles are still sparse. The case without electric field has been studied together with numerical simulations in Ref.~\cite{Sakashita2014}. One can find a large literature on unconfined vesicles in spatially uniform electric fields. DC pulses can, for instance, lead to elongation \cite{Dimova2007}, wrinkling \cite{Knorr2010} or even burst \cite{Riske2009} of the vesicle for strong electric fields. We are not aware of an experimental study of the confined system, but we hope that our results arouse interest in this subject.

In our work we have focused on the effects of polarization and assumed free charges to be absent. These would have to be included to get closer to real experimental conditions, where charges accumulate near the membrane \cite{Dimova2007,Petia2015B}. 
A dynamic study could be very fertile as well given experimental observations that show dynamic transitions of membranes induced by electric fields \cite{Petia2015B,Petia2015A}.


\vspace*{0.5cm}
The authors would like to thank Fabien Pascale for his help on some programming issues. 
The clusters of the PMMS and the LPCT are acknowledged for providing the computer time.


\end{document}


\bibliographystyle{plain}

\title{\LARGE\centering Supplementary material of the article ``Flexoelectric fluid membrane vesicles in spherical confinement'': A brief overview of the numerical methods}

\author{N. Abtahi et al.}
\date{}
\maketitle

\vspace*{-0.7cm}
In this supplementary material we briefly present the numerical methods we have used in our work. The details of both methods \textit{without} the flexoelectric contribution can be found in the Appendices of Ref.~\cite{E19}. Flexoelectricity appends additional terms to the differential equations of the system.

\section{Finite Element Analysis}

\begin{figure}[b]  
\centering
\includegraphics[scale=.8]{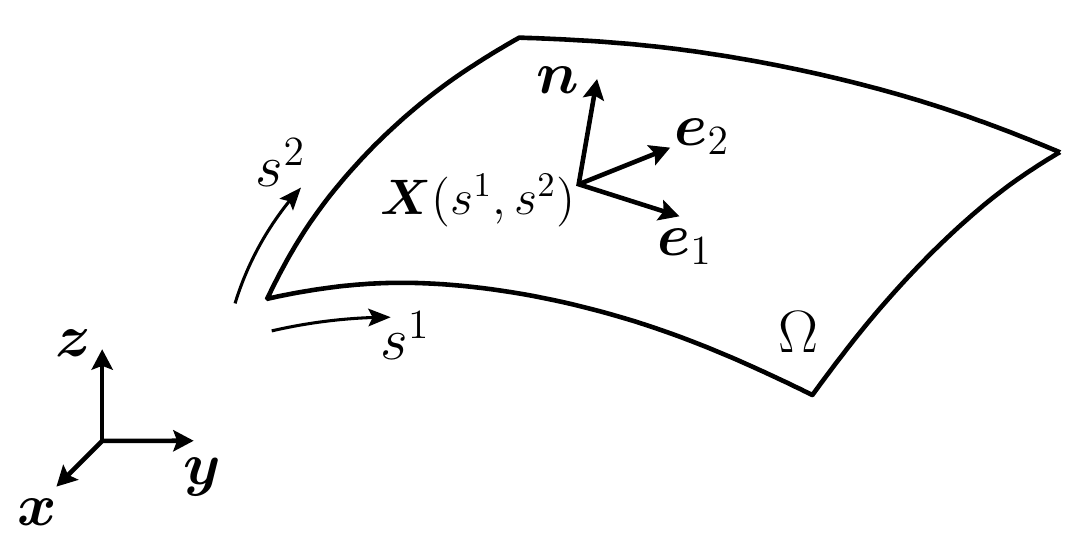}
\caption{Surface parametrization.}
\label{fig:1}
\end{figure}

To find the equilibrium shapes of spherically confined flexoelectric fluid membrane vesicles in an external uniform electric field, we use a finite elements method based on the subdivision surface concept \cite{Cirak2000,Klug2006,Klug2008}. This ensures the necessary $C^1$ continuity of the surface vector function $\VECX(s^1,s^2)$ of the flexoelectric membrane.  
The membrane is parametrized by local curvilinear coordinates $(s^{1},s^{2})$ (see Fig.~\ref{fig:1}). The covariant tangent vectors are given by $\VECe_{\alpha}=\partial_{\alpha}\VECX$ with $\alpha=1,2$. The normal vector is defined as $\VECn=\frac{\VECe_{1}\times\VECe_{2}}{\sqrt{g}}$, and the surface Jacobian is constructed as $\sqrt{g}=\left|\VECe_{1}\times\VECe_{2}\right|$ so that the area $A=\int_{\Omega} \romd A = \int_{\Omega} \sqrt{g} \, \romd s^{1} \romd s^{2}$. 

In the simulations the membrane surface is discretised by a set of triangles with $N\sim 3000$ nodes.
%
The position of the surface is then interpolated by the weighted sum: 
\begin{equation}
\label{eq.3.26}
\VECX_{h}(s^{1},s^{2})=\sum_{i=1}^{N}\VECX_{i}N^{i}(s^{1},s^{2})
\; ,
\end{equation}
where $\VECX_{i}$ is the position of node $i$, and the $N^{i}$ are the loop subdivision trial functions \cite{Cirak2000}.\\
The energy functional 
\begin{equation}
\label{eq:energyFE}
  E^M  = \int_{\Omega} \left\{ 2 \kappa H^{2}+\frac{1}{2D}\left[(\VECE\cdot\VECn)^{2}-\left|\VECE\right|^{2}\right]  \right\} \sqrt{g}  \romd s^{1} \romd s^{2}  \\
  + \int_{\Omega} \frac{\mu_{A}}{2}(\sqrt{g}-\sqrt{\bar{g}})^{2} \, \romd s^{1} \romd s^{2}
    +\frac{\mu_{V}}{2}(V-\bar{V})^{2}
 \end{equation}
is minimised while taking into account the container constraint and self-contacts of the membrane (see below). The constraints on surface and volume are enforced in Eq.~(\ref{eq:energyFE}) \textit{via} a penalty method to improve convergence \cite{E19}. The constants $\mu_{A}$ and $\mu_{V}$ are chosen large enough to fulfill these  constraints to a numerical error of about $10^{-3}$. Their values range from $10^4$ to $10^7$ depending on the value of the electric field $\VECE$. Note that the area constraint is local to avoid mesh distortions during the simulation. 

Energy~(\ref{eq:energyFE}) can be discretised in terms of the nodal positions $\VECX_{i}$. To calculate the force $\VECf_{i}^{M}$ that acts on each node, one considers the variation of the energy with respect to a displacement of the nodal position:
\begin{equation}
\label{eq:variationEM}
\delta{E^M}=\frac{\partial{E^M}}{\partial{\VECX}^{i}}\delta{\VECX}^{i}= -\VECf_{i}^{M}\delta{\VECX}^{i}
\; ,
\end{equation}
which gives:
\begin{equation}
\label{eq:nodalforce}
\VECf_{i}^{M}=\int_{\Omega}{\left[\VECs^{\alpha}\cdot{\partial_{i}\VECe_{\alpha}}+ \VECm^{\alpha}\cdot(\partial_{i}{\VECn})_{,\alpha} 
+\VECf N^{i}
\right]\sqrt{g} \, \romd s^{1} \romd s^{2}}
\; ,
\end{equation}
where 
\begin{eqnarray}
\label{eq:s}
  {\VECs}^{\alpha} & = & 2 \kappa H g^{\alpha\beta}\VECn_{,\beta}+ 2\kappa H^{2}\VECe^{\alpha}  
  -\frac{1}{D}(\VECE\cdot\VECn)(\VECE\cdot\VECe^{\alpha})\VECn-\frac{1}{2D}\left[\VECE^{2}-(\VECE\cdot\VECn)^{2}\right]\VECe^{\alpha} \nonumber \\
  && + \;\; \mu_{A} (\sqrt{g}-\sqrt{\bar{g}}) \VECe^{\alpha}+\mu_{V}\frac{V-\bar{V}}{3}\left[(\VECX\cdot\VECn)\VECe^{\alpha}-(\VECX\cdot \VECe^{\alpha})\VECn\right]
  \; ,
\\
  {\VECm}^{\alpha} & = & -2 \kappa H \VECe^{\alpha} \; , \qquad \text{and}
  \\
  {\VECf} & = & \mu_{V}\frac{V-\bar{V}}{3}{\VECn}
  \; .
\end{eqnarray}
are the stress and moment resultants. In addition to the terms already implemented by Klug and coworkers \cite{Klug2006,Klug2008}, $\VECs^\alpha$ now contains a contribution due to the flexoelectric effect. Note that its continuous version was found in Ref.~\cite{Steigmann2016}.

The contact between the membrane and the container is modeled by a quadratic
force field applied to the nodes that leave the container. The corresponding force is given by
\begin{equation}
\VECf_{i}^{C_1}= - k_{1} d_{i}^{2}\VECn
\; ,
\end{equation}
where $k_{1}\sim 10^5$ is the stiffness constant, $d_{i}$ the penetration depth, and $\VECn$ the outward-pointing normal of the container surface. 
An additional force field is needed to disentangle intersections of the membrane surface that occur during the course of the simulation: at first the contour line of the intersecting polygons is determined. In a second step, the gradient $\VECG_i$ of the contour length with respect to the position of each node $i$ of a triangle involved in the intersection is calculated. The resulting contact force is linear in $\VECG_i$: 
\begin{equation}
\VECf_{i}^{C_2}= k_{2} \VECG_i
\; ,
\end{equation}
where we set $k_2=10$. 

Adding all terms together one obtains the resulting force on each node 
\begin{equation}
\VECf_{i} = \VECf^{M}_{i} + \VECf_{i}^{C_1} + \VECf_{i}^{C_2}
\; .
\end{equation}
During the simulation, we integrate these nodal forces in time according 
to Newton's equations of motion with a damping term until we reach equilibrium, $\VECf_i = 0$
(for more details see Appendix B of Ref.~\cite{E19}).


\section{Hamiltonian formulation for the axisymmetric case}

\begin{figure}
\centering 
   \includegraphics[scale=0.8]{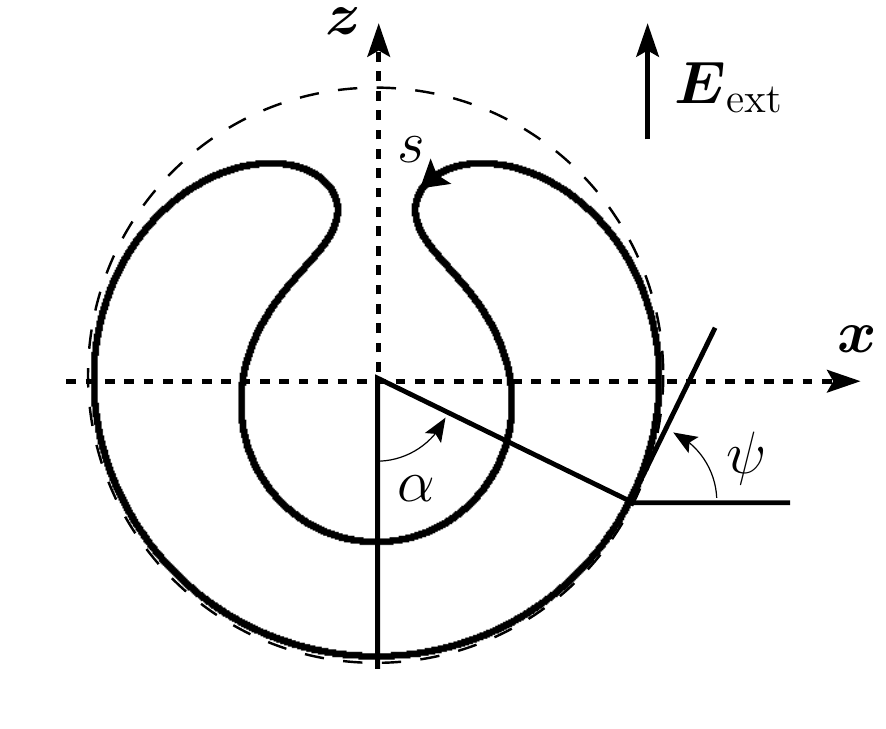}
   \caption{Parametrization of the axisymmetric flexoelectric fluid membrane in spherical confinement. The solid black line represents the cross-section of the membrane. 
   The dashed line indicates the spherical confinement. $s$ is the arc length, $\psi$ is the angle between the $\VECx$ axis and the tangent of the flexoelectric membrane and $\alpha$ represents the detachment angle of the flexoelectric membrane from the confinement. We assume that the membrane consists of two segments, a spherical segment in contact with the confinement (bottom part) and an upper free segment whose shape is determined by solving the Hamilton equations~(\ref{eq:Hamiltonequations}) together with
the appropriate boundary conditions~(\ref{eq:boundconds}).
}
\label{fig:2}
\end{figure}

In the special case of an axisymmetric configuration, one can resort to a Hamiltonian formulation 
of the problem. In the following we assume that the flexoelectric membrane vesicle is symmetric with respect to the $\VECz$ axis (see Fig.~\ref{fig:2}). 
We make use of the angle-arc length parametrization $\psi({s})$, where $\psi$ is the angle between the tangent vector and the $\VECx$ axis, whereas $s$ denotes the arc length.

In this parametrization the area element $\romd A$ is given by:
\begin{equation}
\label{eq:dA}
dA=2{\pi}{\rho} \, \romd s
\end{equation}
and the volume element $\romd V$:
\begin{equation}
\label{eq:dV}
\romd V={\pi}{\rho}^{2}\sin{\psi} \, \romd s
\; .
\end{equation}
The mean curvature is obtained as:
\begin{equation}
\label{eq:meancurvature}
H = - \frac{1}{2}\left( \dot{\psi}+\frac{\sin{\psi}}{\rho} \right)
\; ,
\end{equation}
where the curvature in the meridian direction is $c_{\perp}=-\dot{\psi}$, and the curvature along the parallel direction is $c_{\parallel}=-\frac{\sin{\psi}}{\rho}$. The flexoelectric energy term simplifies to
\begin{equation}
\label{eq:flexoterm}
\frac{1}{2D}\left[(\VECE \cdot\VECn)^{2} - | \VECE |^{2} \right]=\frac{-{\E}^{2}}{2D}\sin^{2}\psi
\; .
\end{equation}
The scaled energy functional of the free part of the membrane vesicle follows as:
\begin{equation}
\label{eq:scaledenergy}
\begin{split}
\tilde{E}& := \frac{E}{\pi\kappa} = \int_{\underline{s}}^{\bar{s}} \tilde{\mathscr{L}} \, \romd s\\
          & = \int_{\underline{s}}^{\bar{s}} \left[ \rho \left( \dot{\psi}+\frac{\sin{\psi}}{\rho} \right)^{2} + 2{{{\tilde{\sigma}}}\rho}
+\lambda_{\rho}(\dot{\rho}-\cos{\psi}) + \lambda_{z}( \dot{z}-\sin{\psi} ) + \tilde{P}\rho^{2}\sin{\psi} - {\rho \Ebar}\sin^{2}{\psi} \right] \romd s ,
\end{split}          
\end{equation}
where $\tilde{\sigma}=\frac{\sigma R^2}{\kappa}$ and $\tilde{P}=\frac{P R^3}{\kappa}$ are the scaled surface tension and pressure, respectively. The Lagrange multiplier functions $\lambda_{\rho}$ and  $\lambda_{z}$ fix the geometrical constraints $\dot{z}=\sin{\psi}$ and $\dot{\rho}=\cos{\psi}$ along the profile. In the integral, $\underline{s}$ represents the arc length at the contact point and $\bar{s}$ corresponds to the arc length at the $\VECz$ axis where the tangent of the membrane is parallel to $\VECx$. 
All lengths are scaled with the radius $R$ of the confining sphere. 
The dimensionless electric field parameter $e$ is defined as
\begin{equation}
  \label{eqn:Ebar}
  \Ebar = \frac{\E^2 R^2}{D \kappa} \; .
\end{equation}

The conjugate momenta of the system are:
\begin{subequations}
\label{eq.4.9m}
\begin{eqnarray}
\label{eq.4.9}
p_{\psi} &=& \frac{\partial{\tilde{L}}}{\partial{\dot{\psi}}}=2\rho\left(\dot{\psi}+\frac{\sin{\psi}}{\rho}\right) \; ,\\
p_{\rho} &=& \frac{\partial{\tilde{L}}}{\partial{\dot{\rho}}}=\lambda_{\rho} \;,\\
p_{z} &=& \frac{\partial{\tilde{L}}}{\partial{\dot{z}}}=\lambda_{z} \; .
\end{eqnarray}
\end{subequations}
One obtains the scaled Hamiltonian \textit{via} the Hamiltonian formalism:
\begin{equation}
\label{eq.4.10}
\begin{split}
\tilde{\mathscr{H}} & = \dot{\psi}p_{\psi}+\dot{\rho}p_{\rho}+\dot{z}p_{z}-\tilde{\mathscr{L}}\\
      & = \frac{p_{\psi}^{2}}{4\rho}-p_{\psi}\frac{\sin\psi}{\rho}-2{{{\tilde{\sigma}}}\rho}+p_{\rho}\cos\psi+p_{z}\sin{\psi}-\tilde{P}\rho^{2}\sin{\psi}+
\Ebar {\rho}\sin^{2}\psi
\end{split}
\end{equation}
yielding the Hamilton equations:
\begin{subequations}
\label{eq:Hamiltonequations}
\begin{eqnarray}
\dot{\psi} &=& \frac{\partial{\tilde{\mathscr{H}}}}{\partial{ p_{\psi}}} =\frac{p_{\psi}}{2\rho}-\frac{\sin{\psi}}{\rho}\; ,\\
\dot{\rho} &=& \frac{\partial{\tilde{\mathscr{H}}}}{\partial p_{\rho}} =\cos{\psi} \; ,\\
\dot{z} &=& \frac{\partial{\tilde{\mathscr{H}}}}{\partial p_{z}} =\sin{\psi}\; ,\\
\dot{p}_{\psi} &=& -\frac{\partial{\tilde{\mathscr{H}}}}{\partial{\psi}} =\left(\frac{p_{\psi}}{\rho}+\tilde{P}\rho^{2}-p_{z}\right)\cos{\psi}+p_{\rho}\sin{\psi}-2\Ebar{\rho}{\sin{\psi}}{\cos{\psi}} \; ,\\
\dot{p}_{\rho} &=& -\frac{\partial{\tilde{\mathscr{H}}}}{\partial{\rho}} =\frac{p_{\psi}}{\rho}\left(\frac{p_{\psi}}{4\rho}-\frac{\sin{\psi}}{\rho}\right)+2\tilde{\sigma}+2\tilde{P}\rho\sin{\psi}-\Ebar{\sin^{2}{\psi}} \; ,\\
\dot{p}_{z} &=& -\frac{\partial{\tilde{\mathscr{H}}}}{\partial{z}}= 0 \; .
\end{eqnarray}
\end{subequations}
For $e=0$ one obtains the classical Hamilton equations of a lipid membrane vesicle as expected \cite{E19}. The flexoelectric effect adds terms which are linear in $\Ebar$ and simple analytical functions of the surface parametrization. The Hamilton equations can be solved with a standard shooting method \cite{B32} subject to boundary conditions which we discuss in the following.

The Hamiltonian $\mathscr{H}$ does not explicitly depend on the arc length $s$. Since we have not fixed the total arc length $\bar{s}-\underline{s}$ for the integration, the Hamiltonian is conserved: 
\begin{subequations}
\label{eq:boundconds}
\begin{equation}
\label{eq:conservationofH}
\tilde{\mathscr{H}}=0\; .
\end{equation}
At the contact point ($s=\underline{s}$) the free part of the flexoelectric membrane detaches from the container and the angle $\psi$ has to equal $\alpha$ since the membrane must not have kinks. At the $\VECz$ axis ($s=\bar{s}$) the free profile is horizontal, which leaves us with the following boundary conditions:
\begin{equation}
\label{eq:boundcondangles}
\psi(\underline{s})=\alpha \; , \qquad \psi(\bar{s})=\pi \; .
\end{equation}
A variation of the contact line as was, for example, done for the non-electric case in Refs.~\cite{Deserno2007,Steigmann2009}, yields:
\begin{equation}
\label{eq:contactcurvcond}
\dot{\psi}(\underline s)=1+\sqrt{\left|\Ebar_{\text{c}}-\Ebar_{\text{f}}\right|}\sin\alpha\; ,
\end{equation}
\end{subequations}
where $\Ebar_{\text{c}}$ is the electric field parameter at the membrane in contact with the confinement and $\Ebar_{\text{f}}$ is the electric field parameter of the free membrane.
For a uniform external electric field, however, the second term equals zero and we are left with the 
classical contact curvature condition of the case without electric field. 

The Hamilton equations are integrated with a fourth-order Runge-Kutta method. For a fixed $\tilde{\sigma}$ and $\tilde{P}$ and a trial angle $\alpha$ we search for shapes which fulfill all of the boundary conditions. When a profile is found we calculate its area and volume \textit{a posteriori}. By scanning the parameter space ($\tilde{\sigma},\tilde{P}$) we obtain vesicles with variable area and volume. 
For more details we again refer to Ref.~\cite{E19}.


\bibliography{suppbib}